\documentclass[nofootinbib,prd,superscriptaddress]{revtex4}

\usepackage{amstext,amsmath,amssymb,amsfonts,bbm}
\usepackage{amsthm}
\usepackage[dvips]{graphicx}
\usepackage{latexsym}
\usepackage{euscript}
\usepackage{epsfig}
\usepackage{psfrag}
\usepackage{ulem}
\usepackage{color}

\def\be{\begin{equation}}
\def\ee{\end{equation}}
\def\bes{\begin{eqnarray}}
\def\ees{\end{eqnarray}}

\def\arr{\rightarrow}

\def\w{\wedge}
\def\la{\langle}
\def\ra{\rangle}
\def\6{\langle}
\def\9{\rangle}
\def\half{\mbox{$\f 1 2$}{}}
\def\tree{\mbox{$\f 3 2$}{}}
\def\f{\frac}
\def\tl{\widetilde}
\def\tg{\tl{g}}
\def\tG{\tl{G}}\def\tB{\tl{B}}\def\tH{\tl{H}}
\def\tC{\tl{C}}

\def\cH{{\cal H}}
\def\cO{{\cal O}}

\DeclareMathOperator{\tr}{tr}

\newtheorem{result}{Result}

\def\Id{{\mathbbm 1}}
\def\1{{\mathbbm 1}}

\def\cL{{\cal L}}

\def\aa{{\mathbf{A}}}
\def\bb{{\mathbf{B}}}

\newcommand{\N}{\mathbb{N}}
\newcommand{\Z}{\mathbb{Z}}

\newcommand{\R}{\mathbb{R}}

\def\v{\vec}
\newcommand{\SU}{\mathrm{SU}}

\newcommand{\Ref}[1]{(\ref{#1})}

\def\ext{{\rm ext}}
\def\inte{{\rm int}}
\def\rb{{\rm b}}
\def\rf{{\rm f}}
\def\out{{\rm out}}

\def\sg{\mbox{\sl g}}

\begin{document}
\title{The  entropic boundary law in BF theory}
\author{ Etera R. Livine}\email{etera.livine@ens-lyon.fr}
\affiliation{Laboratoire de Physique, ENS Lyon, CNRS UMR 5672, 46 All\'ee d'Italie,
69007 Lyon, France EU}
\author{ Daniel R. Terno}\email{dterno@physics.mq.edu.au}
\affiliation{Centre for Quantum Computer Technology, 
Macquarie University, Sydney NSW 2109, Australia}

\begin{abstract}

We compute the entropy of a closed bounded region of space for pure 3d Riemannian gravity
formulated as a topological BF theory for the gauge group $\SU(2)$ and show its holographic
behavior. More precisely, we consider a fixed graph embedded in space and study the flat connection
spin network state without and with particle-like topological defects. We regularize and compute
exactly the entanglement for a bipartite splitting of the graph and show it scales at leading order
with the number of vertices on the boundary (or equivalently with the number of loops crossing the
boundary). More generally these results apply to BF theory with any compact gauge group in any
space-time dimension.

\end{abstract}

\maketitle


\section*{Introduction}

Entropy is a key notion in the search for a quantum gravity, both as the thermodynamic quantity
useful to probe the physics and potential phenomenology of the theory and as the measure of
information useful to identify the physical degrees of freedom and their dynamics. Research has
focused on the particular case of black holes and has lead to the holographic principle directly
relating geometric quantities (the area) to the entropy in quantum gravity.

In the context of Loop Quantum Gravity (see \cite{lqg} for reviews), most of the black hole entropy
calculations have been performed in the framework of isolated horizons following the seminal work
by Ashtekar, Baez and Krasnov \cite{bhlqg}. Assuming precise boundary conditions for the
gravitational fields on the horizon, they count the number of (kinematical) boundary states
consistent with fixing the value of the area. Here we would like to get rid of the classical
boundary: instead of considering a spin network state with a boundary specified classically, we
look at a closed spin network and an arbitrary bipartite splitting into inside/outside regions (see
e.g \cite{us:loops}). The aim is to compute the entanglement between these two parts of the spin
network for a physical quantum geometry state solving the Hamiltonian constraint. The first step
that we take here is to work this out for BF theory instead of gravity. Indeed BF theories are
topological field theories lacking local degrees of freedom, thus allowing for an exact
quantization (see e.g \cite{baez}). In this context, we know the physical states solving all the
constraints and can compute the entanglement explicitly. This turns out to be similar to ground
state entanglement calculations in some spin models developed for topological quantum computation
\cite{kitaev,hamma}.

The motivation to analyze BF theory is that it is very close to gravity. First, in three space-time
dimensions, 3d gravity is actually a topological BF theory with the Lorentz group as gauge group.
Second, in four space-time dimensions, gravity can be reformulated as an constrained BF theory and
we can work on a quantization scheme with quantum BF theory as the starting point. Of course, the
fact that gravity has local degrees of freedom should matter in the end. Nevertheless, studying BF
theory should allow us to develop mathematical tools and procedures later useful for loop gravity.

In the present paper, we start with a quick overview of BF theory and the definition of the
physical quantum states as spin network states for the flat connection with possibly particle-like
topological defects. Then focusing on  $\SU(2)$ BF theory, we explicitly compute the entanglement
between the two parts of such a flat spin network states and we show its holographic behavior: it
scales with the size of the boundary (more precisely, with the number of boundary vertices). We
also study the influence of topological defects. We show that they do only affect the entropy when
located on the boundary between the two regions and we compute the finite variation of entanglement
that they create.

\section{An Overview of BF Theory}

\subsection{$\SU(2)$ BF theory: spin networks and physical states}

BF theory is a class of topological gauge field theories defined on
a oriented smooth $n$-dimensional manifold ${\cal M}$ by the
following action (see e.g. \cite{baez} for a review):
\be
S[A,B]\,=\,\int_{{\cal M}}\tr(B\w F[A]).
\ee
The gauge group is a (semi-simple compact) Lie group $G$ whose Lie algebra $\mathfrak{g}$ is
equipped with an invariant (non-degenerate) bilinear form $\tr(\cdot,\cdot)$.
Picking a local trivialization of the principal $G$-bundle over $\cal M$, the basic fields are a
$\mathfrak{g}$-valued connection 1-form $A$ with curvature 2-form $F[A]$ and a
$\mathfrak{g}$-valued ($n$-2)-form $B$. The action is invariant under the action of the gauge
group, for $h\in G$:
\be
B\arr hBh^{-1},
\qquad
A\arr hAh^{-1} + hdh^{-1}.
\label{Gsym}
\ee
It is also invariant under shifts of the $B$ field by an arbitrary (n-3)-form $\phi$:
\be
B\arr B + d_A \phi,\qquad \delta A\,=0.
\label{toposym}
\ee
This symmetry kills all local degree of freedom making BF theory a topological field theory. Its
classical field equation impose a flat connection, $F[A]=0$, and a vanishing ``torsion", $d_A B
=0$. In particular, in $n=3$ space-time dimensions, BF theory for the gauge groups $G=\SU(2)$ and
$G=\SU(1,1)$ is equivalent to 3d Riemannian and Lorentzian gravity (in its first order
formulation).

For our present purpose, we are interested into BF theory from the
point of view of its canonical quantization {\it \`a la} loop
quantum gravity. In the following, we will work with $n=3$ and
$G=\SU(2)$ although all the formalism and techniques apply to
arbitrary space-time dimensions and arbitrary (compact semi-simple)
Lie groups. We perform a 2+1 splitting of the 3d space-time viewing
the 3d manifold ${\cal M}\sim \Sigma \times \R$ as a two-dimensional
space $\Sigma$ evolving in time. Then the spatial parts of the
connection $A$ and the field $B$ on $\Sigma$ are canonically
conjugate variables. The theory is completely constrained and the
Hamiltonian vanishes on-shell. There are two sets of constraints,
 imposed by the time components of $A$ and $B$, respectively, which
turn out to be simply Lagrange multipliers. The first constraints
impose the flatness of the connection on $\Sigma$. The second set of
constraints imposes the vanishing of the torsion on $\Sigma$ and is
usually called the {\it Gauss law}. These are first class
constraints, respectively generating the translational symmetry
\Ref{toposym} and the gauge invariance \Ref{Gsym} under the action
of the group $G$. For more details on the canonical analysis and
resulting structures, the interested reader can check \cite{lqg}.

The loop quantization scheme is based on a specific choice of wave
functions. We choose cylindrical functionals of the connection $A$
on $\Sigma$. More precisely, they depend on $A$ through only a
finite number of variables: they are functions
of the holonomies of $A$ along the edges of some arbitrary (finite)
oriented graph in $\Sigma$. Considering a particular graph $\Gamma$
with $E$ and $V$ vertices, we consider the holonomies
$g_1[A],..,g_E[A]$ of the connection along the edges $e=1,..,E$ of
$\Gamma$ and build wave functions of the following type:
$$
\psi_\Gamma(A)\,=\,
\psi(g_{1}[A],..,g_{E}[A]).
$$
Further we require these functionals to be gauge-invariant. Since the action of the group on the
connection translates into a group action at the end points of the holonomies, $g_e[A]\arr
h_{s(e)}^{-1}g_e[A] h_{t(e)}$, where $s(e)$ and $t(e)$ are respectively the source and target
vertices of the edge $e$, the gauge invariance of the wave functions $\psi_\Gamma$ involves a
invariance under the group action at every vertex of the graph $\Gamma$:
\be
\forall h_v\in G^{\times V},\quad
\psi(g_{1},..,g_{E})\,=\,
\psi(h_{s(1)}^{-1}g_{1}h_{t(1)},..,h_{s(E)}^{-1}g_{E}h_{t(E)}).
\label{gaugeinv}
\ee

Then we need to impose the flatness condition $F=0$ on these wave functions. To keep the discussion
as simple as possible, we consider a graph $\Gamma$ which forms a lattice faithfully representing
the canonical surface $\Sigma$. More precisely, if we cut the surface $\Sigma$ along the embedded
graph $\Gamma$, then we are left with surfaces all homomorphic to the unit disk. These are the
faces of the lattice. We now impose that the holonomy of the connection around each face is the
identity. Thus the projection onto physical states satisfying the flatness condition is implemented
by the multiplication by $\delta$-function around each face:
\be
\psi(g_e)\,\mapsto\,
\prod_{\cal L} \delta\left(\prod_{e\in {\cal L}} g_e\right)\,\psi(g_e),
\ee
where ${\cal L}$ labels the loops around the faces of the lattice. This imposes trivial holonomies
around every contractible loop in $\Sigma$ while allowing for arbitrary holonomies around the
non-contractible cycles of the surface.

When $\Sigma$ is the two-sphere, this gives a single flat quantum state. When $\Sigma$ is an
orientable compact surface of genus $n$, the physical space of flat quantum states is isomorphic to
the space of $L^2$ functions of $2n$ group elements $A_1,B_1,..,A_{n},B_{n}$ satisfying
$A_1B_1A_1^{-1}B_1^{-1}..A_{n}B_{n}A_{n}^{-1}B_{n}^{-1}=\mathbbm{1}$ and invariant under diagonal
conjugation $A_i,B_i\,\mapsto hA_ih^{-1},hB_ih^{-1}$ for $h\in G$. For instance, for the two-torus,
physical states will be gauge invariant functions of two group elements $A$ and $B$ satisfying
$ABA^{-1}B^{-1}=\mathbbm{1}$.

Finally, in 2+1 spacetime dimensions particles create conical
singularities, which leave the spacetime flat except along their
worldline. They are represented as topological defects, which
translate into non-trivial holonomies. For a spinless particle on a
given face of the lattice, we replace the $\delta$-function by a
$\delta_\theta$-function imposing that the holonomy around the face
has a class angle $\theta$. This deficit angle $\theta$ then defines
the mass of the particle (e.g. see
\cite{grav,frei} for more details).

Our goal in the present work is to compute the entanglement on a physical state between a region of
$\Sigma$ and the rest of the surface and to check whether it satisfies to an ``area-entropy" law.
On a given graph $\Gamma$, we will consider a connected region of the graph and compute the von
Neumann entropy of the reduced density matrix obtained after tracing out all the holonomies outside
that region from a given physical state. In the next part, we will show how this works for a BF
theory based on a discrete group, in which case the theory can be reformulated as a spin system.

To conclude this review section, we give a few mathematical details on the $\SU(2)$ group. We
parameterize group elements $g$ as
\be
g=
e^{i\theta\hat{u}.\v{\sigma}} =\cos\theta\,\Id+i \sin\theta\,\hat{u}.\vec{\sigma},
\ee
where $\theta\in[0,2\pi]$ is the class angle (or half of the rotation angle), $\hat{u}$ is a unit
vector on the 2-sphere indicating the axis of the rotation and $\v{\sigma}$ are the standard Pauli
matrices. Let us point out $g(\theta,\hat{u})=g(-\theta,-\hat{u})$. In these variables the
normalized Haar measure $dg$ on $\SU(2)$ reads as
\be
\int_{\SU(2)}dg\,f(g)=\,
\f1{2\pi^2}\int_{0}^{\pi}\sin^2\theta\,d\theta\,
\int_{{\cal{S}}^2}d^2\hat{u}\,
f(\theta,\hat{u}).
\ee
By the Peter-Weyl theorem, every function which is invariant under
conjugation can be decomposed on the characters of the irreducible
(spin) representations of $\SU(2)$. Such representations are labeled
by a half-integer $j\in\N/2$ and the corresponding characters are
\be
\chi_j(g)\,=\,\tr_j\left(D^j(g)\right)
=\f{\sin(2j+1)\theta}{\sin\theta}=U_{2j}(\cos\theta),
\ee
where $U_n$ is the $n$-th Tchebychev polynomial of the second kind. The $\delta$-distribution
decomposes as:
$$
\delta(g)=\sum_{j\in\N/2}(2j+1)\chi_j(g).
$$
Finally, we introduce the distributions $\delta_\theta$ that localize group elements on a specific
equivalence class under conjugation by fixing their rotation angle:
\be
\int dg\, \delta_\theta(g) f(g)
=\f{1}{4\pi}\int_{{\cal{S}}^2}d^2\hat{u} f((\theta,\hat{u})).
\ee
Its decomposition into characters reads:
\be
\delta_\theta(g) = \sum_{j\in\N/2}\chi_j(\theta)\chi_j(g).
\label{deltatheta}
\ee

\subsection{A word on Kitaev's spin system and $\Z_2$ BF theory}

In this subsection, we give a quick overview of Kitaev's spin system introduced in \cite{kitaev}
and the corresponding ground state entanglement calculations presented in \cite{hamma}. As in the
previous section, we consider the (canonical) surface $\Sigma$ provided with the lattice defined by
the graph $\Gamma$. We attach a two-level system (which is called a qubit in the language of
quantum information and represents a spin-$1/2$ particle) to each edge $e$ of the graph. Label its
basis  states as $|\pm_e\ra$. Consider the following Hamiltonian which is a sum of local operators
attached to the vertices and the faces of the lattice:
\be
H\,=\, -\sum_v \bigotimes_{e\ni v}\sigma_x^{(e)} -\sum_f \bigotimes_{e\in f}\sigma_z^{(e)},
\ee
with $\sigma_x \,|\pm\ra =|\mp\ra$ and $\sigma_z \,|\pm\ra =\pm|\pm\ra$. Calling the operators
$\aa_v\equiv\otimes_{e\ni v}\sigma_x^{(e)}$ and $\bb_f\equiv\otimes_{e\in f}\sigma_z^{(e)}$, we
first notice that all these operators commute with each other. We can thus diagonalize them
simultaneously. Ground states $|\psi_0\ra$ are then states which diagonalize all $\aa_v$ and
$\bb_f$ with the highest eigenvalue:
\be
\aa_v|\psi_0\ra\,=\,\bb_f|\psi_0\ra\,=\,|\psi_0\ra.
\ee
As shown in \cite{kitaev} (and reviewed in \cite{hamma}), this spin system is equivalent to BF
theory based on the discrete gauge group $\Z_2$. Generalizing this to systems with a higher number
of levels attached to each edge allows to reformulate in a similar fashion BF theory for an
arbitrary discrete group.

Ground states correspond to physical states in BF theory. The holonomy can either be $+$ or $-$
along an edge. Then the $\aa_v|\psi_0\ra\,=\,|\psi_0\ra$ condition implements the Gauss law
imposing gauge invariance at each vertex, while the face condition $\bb_f|\psi_0\ra\,=\,|\psi_0\ra$
imposes the flatness of the $Z_2$ holonomy. Thus the Hilbert space of ground states for an
orientable surface of genus $g$ has a $2^{2g}$ degeneracy, with the holonomy around all $2g$ cycles
of the surfaces being free.

Finally, using the stabilizer space methods, the entropy of one (bounded connected) region of the
lattice was computed in any ground state \cite{hamma}. It was shown that it does not depend on the
particular ground state (i.e non-contractible cycles do not matter) and it is simply related to the
perimeter of the region's boundary as
$$
S=n_L-1,
$$
where $n_L$ is the number of spins in the perimeter of a region (for a 2d square lattice).

In the present work, we generalize these entanglement calculation to the case of the continuous Lie
group, making explicit calculations in the case of the group $G=\SU(2)$. We do not use the same
methods as developed in the analysis of the spin systems but exploit simple loop quantum gravity
tools.

\section{Entanglement: generic structures and Entropy-Boundary law}

\subsection{The setting: flat spin networks and choice of independent loops}

Let us consider a fixed connected oriented (abstract) graph $\Gamma$ and a spin network wave
functional $\psi(g_e)$ living on it. We would like to study the completely flat wave functional,
imposing that the holonomy along all loops of the graph. This corresponds to the flat physical
state for BF theory for a trivial topology (a two-sphere) of the canonical surface $\Sigma$. It
also gives the physical state for a non-trivial topology but with the additional constraint of
trivial holonomies around every cycle of the canonical surface. To truly consider non-trivial
topologies, we would need to alow for non-trivial holonomies around some of the loops of the graph
(which would then represent the non-contractible cycles of the canonical surface).

In order to study the entanglement properties of this completely flat state, we write the flat wave
functional $\psi_0(g_e)$ as a product of $\delta$-functions on the group $\SU(2)$ for every loop of
the graph. However, such a naive product would obviously lead to infinities due to redundant
$\delta$-functions and we need to consider the product over independent loops. We formalize this
using by the gauge fixing procedure for spin networks that was developed in \cite{trees}.

Considering a generic spin network functional $\psi(g_e)$ on
$\Gamma$, we can gauge fix the $\SU(2)$ gauge transformations
\Ref{gaugeinv} acting at the vertices by introducing a
maximal tree $T$ on the graph. $T$ is a connected set of edges on $\Gamma$ touching all vertices of
$\Gamma$ but never forming any loop. If we call $E$ and $V$ respectively the number of edges and
vertices of $\Gamma$, then the number of edges in $T$ is exactly $V-1$. On the other hand, the
number of edges not in $T$ is exactly the number of independent loops on the graph
$L\,\equiv\,E-V+1$. The gauge fixing consists in fixing all the group elements $g_e$ living on
edges $e\in T$ belonging to the tree to the identity. More precisely, we choose a reference vertex
$v_0$. Then for all vertices $v$, there exists a unique path $[v_0\arr v]$ linking $v_0$ to $v$
along the tree $T$. We perform a gauge transformation with parameters:
$$
h_v \,\equiv\, \underset{{e\in [v_0\arr v]}}{\overrightarrow{\prod}} g_e.
$$
For all edges on the tree $e\in T$, the resulting group elements $h_{s(e)} g_e h_{t(e)}^{-1}$ are
set to the identity $\Id$. For all edges not belonging to the tree $e\notin T$, this defines a loop
variable $G_e \equiv\, h_{s(e)} g_e h_{t(e)}^{-1}$, which  is the oriented product of the loop
$\cL_e\,\equiv\,[v_0\arr s(e)]\cup e\cup [t(e)\arr v_0]$ starting at $v_0$ and going to $s(e)$
along the tree $T$, and then coming back to $v_0$ along the tree from the the vertex $t(e)$. The
procedure ensures that (see \cite{trees} for more details) the wave functions evaluated on the
original $g_e$'s is equal to its evaluation on the $G_{e\notin T}$ while setting the other group
elements to $\Id$.

Finally, we define the completely flat spin network state as:
\be
\psi_0(g_e)\,\equiv\,\prod_{e\notin T}\delta(G_e).
\ee
It is straightforward to check that the resulting state actually does not depend on the choice of
neither the reference vertex $v_0$ nor the maximal tree $T$. Moreover it truly imposes the
condition that the holonomy around any loop of the graph $\Gamma$ is constrained to be equal to the
identity $\Id$.

Such a distributional state is obviously not normalisable for the kinematic inner product defined
with the Haar measure $\int dG_e$. It is indeed $L^1$  but not $L^2$. To deal with it, we need to
regularize it. The method we will use is the standard one when working in loop quantum gravity and
spin foam models \cite{lqg}: expanding the state in $\SU(2)$ representations, we will introduce by
hand a cut-off $J$ in the representations and then study the behavior of the various quantities in
the large spin limit $J\arr+\infty$.

\medskip

Our purpose is to consider a bounded region $A$ of the graph $\Gamma$ and compute the entanglement
between $A$ and the rest of the graph on the completely flat spin network state. This will give the
entropy of $A$.
More precisely, we choose a connected region $A$ of $\Gamma$. We define it as a set of $V_\inte$
vertices and the $E_\inte$ edges that link them (i.e any edge whose both source and target vertices
are in $A$ also belongs to $A$). We further distinguish the $E_\rb$ boundary edges, that connect
one vertex inside $A$ to one vertex outside, and the $E_\ext$ exterior edges who do not touch the
considered region $A$. We call the interior graph $\Gamma_\inte$ the graph formed by the vertices
and (interior) edges of $A$.  The exterior graph $\Gamma_\ext$ is defined as its complement: it
consists in both exterior and boundary edges. We define the $V_\rb$ boundary vertices which belong
to both interior and exterior graphs i.e vertices in $A$ that touch some boundary edges. Then the
exterior graph has  $V_\ext=V-V_\inte+V_\rb$ vertices.

To define the entanglement between the interior and the exterior regions, we consider the reduced
density matrix on $A$ obtained by tracing out all the holonomies outside $A$ from the full density
matrix:
\be
\rho(g_e,\tg_e)\,\equiv\,\psi(g_e)\bar{\psi}(\tg_e),\qquad
\rho_\inte(g_{e\in A},\tl{g}_{e\in A})
\,=\,
\int [dg_{e\notin A}]\,
\psi(g_{e\in A},g_{e\notin A})\bar{\psi}(\tl{g}_{e\in A},g_{e\notin A}).
\label{reduced}
\ee
We are then interested in the standard measure of entanglement defined as $E\,=-\tr_A
\rho_\inte\log\rho_\inte$. We would like to compute it on the completely flat spin network state, which a
physical state for BF theory. For this purpose, we need to adapt the choice of the tree $T$ used in
defining the flat state $\psi_0$ to the choice of the studied region $A$: we would like the
definition of $\psi_0$ to respect the interior/boundary/exterior structure.

We now choose two reference vertices $v_0$ in the interior (there is no problem if $v_0$ itself is
a boundary vertex) and $w_0$ in the exterior (for the sake of simplicity, we choose $w_0$ so that
it is not a boundary vertex) and two maximal trees $T_\inte,T_\ext$ respectively on the interior
and outer graphs. We would like to form a maximal tree $T$ on the whole graph by merging the two
trees. The only issue is that the interior and outer graphs, and thus the two trees, both share the
$V_\rb$ boundary vertices. Considering a boundary vertex $v_\rb$, there exists a unique path along
the tree $T_\inte$ from $v_0$ to $v_\rb$ and there also exists a unique path in the exterior along
$T_\ext$ from $w_0$ to $v_\rb$. If we consider the straightforward gluing of the two trees
$T_\inte\cup T_\ext$, then we obtain loops as soon as there are at least two boundary vertices of
the type $[v_0\arr v_\rb^{(1)} \arr w_0 \arr v_\rb^{(2)}\arr v_0]$. In order to get a tree, we
simply need to break these loops. For this purpose, we single out an arbitrary boundary vertex
$v_\rb^{(0)}$ and we number the other boundary vertices $v_\rb^{(i)}$ with $i=1,..,(V_\rb-1)$. We
will remove one edge from each loop $[v_0\arr v_\rb^{(0)} \arr w_0 \arr v_\rb^{(i)}\arr v_0]$ along
$T_\inte\cup T_\ext$. More precisely, for each $v_\rb^{(i)}$, we consider the unique boundary edge
$e^{(i)}$ in $T_\ext$ touching $v_\rb^{(i)}$. Finally, we define $T\,\equiv\,(T_\inte\cup
T_\ext)\setminus
\{e^{(i)}\}$ and it is straightforward to check that $T$ is a maximal tree on the whole graph
$\Gamma$.
Indeed there are no loops in $T$. Moreover there exists a path in $T$ from $v_0$ to any vertex
$v\in\Gamma$: if $v$ is in the region $A$ the path is simply the path $[v_0\arr v]$ within
$T_\inte$ while if $v$ is outside $A$ we consider the sequence of edges $[v_0\arr v_\rb^{(0)} \arr
w]
\cup [w \arr v]$ with the first halves of the path in $T_\inte$ and the second in $T_\ext$.

We follow the previous gauge fixing procedure and define the flat state $\psi_0$ according to this
chosen maximal $T$. We have one holonomy loop variable per edge $e\notin T$ not belonging to the
tree $T$. There are three types of such edges. First, we identify the edges $e$ in the region $A$
but not belonging to the interior tree $T_\inte$. Second, we identify the edges $e$ outside $A$ but
not belonging to the outer tree $T_\ext$. Finally, there are the boundary edges $e^{(i)}$,
$i=1,..,(V_\rb-1)$. For all these edges $e\notin T$, we associate the corresponding holonomies
around the interior, exterior and boundary loops respectively which consist in edges $f$ going from
$v_0$ to $s(e)$ along $T$ then along the edge $e$ then coming back from $t(e)$ to $v_0$ along $T$.
We call the loops $\cL_e\,\equiv [v_0\arr s(e)]\cup e\cup [t(e)\arr v_0]$ and define the
corresponding holonomies:
\be
\forall e\in\Gamma_\inte\setminus T_\inte,
\quad G_e\equiv\underset{{f\in \cL_e}}{\overrightarrow{\prod}} g_f,
\qquad
\forall e\in\Gamma_\ext\setminus T_\ext,
\quad H_e\equiv\underset{{f\in \cL_e}}{\overrightarrow{\prod}} g_f,
\qquad
\forall i=1,..,(V_\rb-1),
\quad B_i\equiv\underset{f\in \cL_{e^{(i)}}}{\overrightarrow{\prod}} g_f.
\ee
The flat state is then the product of $\delta$-functions over all these loops:
\be
\psi_0(g_e)\equiv \prod_{e\in\Gamma_\inte\setminus T_\inte}
\delta(G_e)\,\prod_{e\in\Gamma_\ext\setminus T_\ext}\delta(H_e)\,\prod_{i=1..(V_\rb-1)}\delta(B_i).
\ee
We insist that the whole procedure with the choice of a tree $T$ is simply to choose a set of
independent loops in order to impose the flatness conditions with no redundant
$\delta$-distribution. In other words, the state $\psi_0$ as defined above still imposes that the
holonomy around any loop of the whole graph $\Gamma$ is trivial.

Finally, a last detail is that we can cut the loops $\cL_e$ for exterior edges
$e\in\Gamma_\ext\setminus T_\ext$. We can have them start at the exterior vertex $w_0$ instead of
the reference vertex $v_0$: we define $\tl{\cL}_e\,\equiv[w_0\arr s(e)]\cup e\cup [t(e)\arr w_0]$.
This leads to the holonomies $H^{-1} H_e H$ where $H$ is the oriented product of the group elements
from $v_0$ to $w_0$ along the tree $T$ (going through $v_\rb^{(0)}$). Since the $\delta$-functions
are central, replacing the $H_e$'s by $H^{-1} H_e H$ does not change anything to the definition of
the flat state $\psi_0$. On the other hand, the loops $\tl{\cL}_e$ have the advantage that there
only involve edges in the outer graph $\Gamma_\ext$ (i.e not belonging to the region $A$). We will
therefore use this prescription for the following entanglement calculations.

\begin{figure}[htbp]
\epsfxsize=0.5\textwidth
\centerline{\epsffile{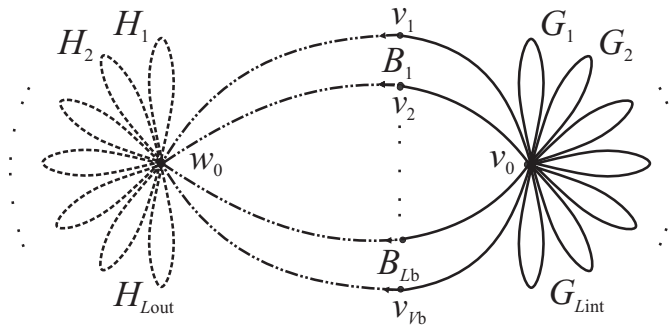}}
\caption{\small The interior and the exterior flowers are linked by $V_\rb$ boundary edges. Interior edges are shown as solid lines.}\label{flowers}
\end{figure}

We would like to underline that the properties (entanglement,..) of the flat state $\psi_0$ do not
depend on the specific choice of tree $T$ that we made. Indeed the choice of tree is simply a
choice to gauge fixing for the spin network functional. The particular tree that we defined using
trees in the interior and exterior regions is a convenient choice allowing a clear description of
the ``boundary loops" which are the central objects for the entanglement calculation.

\subsection{Computing the entanglement for the flat state}

We now compute the reduced density matrix $\rho_\inte$ obtained by
tracing out all exterior holonomies from the flat state $\psi_0$.
The von Neumann entropy of this reduced density matrix defines the
entanglement between the region $A$ and the rest of the spin
network. Our main result is that the entanglement scales with the
size of the boundary:
\be
E_A\equiv S[\rho_\inte] \equiv -\tr_A \rho_\inte \log \rho_\inte \,=\, (V_\rb-1)\,u(J),
\ee
where $(V_\rb-1)$ is the number of boundary loops between the region $A$ and the exterior and
$u(J)$ is a unit of entanglement which only depends on the regulator $J$ (representation cut-off)
and does not depend on either the graph $\Gamma$ or the choice of the region $A$. In particular, if
there is a single boundary vertex, $V_\rb=1$, the region $A$ is totally disentangled from the rest
of the spin network.

Let us compute the reduced density matrix as defined by \Ref{reduced}:
$$
\rho_\inte(g_e,\tg_e)\,=\,
\int [dg_{e\notin A}d\tg_{e\notin A}]\,\prod_{e\notin A}\delta(\tg_eg_e^{-1})\,
\prod_{e\notin T} \delta(G_e)\delta(B_i)\delta(H_e)\,\prod_{e\notin T} \delta(\tG_e)\delta(\tB_i)\delta(\tH_e).
$$
First, the interior loops are unaffected by the integration over exterior edges since they do not
involve any edges outside the region $A$. Second, the exterior loops involve only exterior edges,
thus we have the identification $\tH_e=H_e$. This produces $\delta(\Id)$ infinities which we
re-absorb in the normalisation of the reduced density matrix. Finally, the moot point is what
happens to the boundary loops. More precisely, we have defined:
\be
B_i=  C_0 D_0^{-1} D_i C_i^{-1}, \qquad\textrm{with}
\quad C_i\equiv \prod_{f\in [v_0\arr v_\rb^{(i)}] \subset \Gamma_\inte} g_f,
\quad D_i \equiv\prod_{f\in [w_0 \arr v_\rb^{(i)}] \subset \Gamma_\ext} g_f .
\ee
With this decomposition the identification $\delta(\tg_eg_e^{-1})$ for all edges $e\in\Gamma_\ext$
implies that the factors $\delta(B_i)\delta(\tB_i)$ can be re-written as
$\delta(C_i^{-1}C_0\tC_0^{-1}\tC_i)$ where the holonomy $C_i^{-1}C_0$ is the (oriented) product of
the group elements from the boundary vertex $v_\rb^{(i)}$ to $v_\rb^{(0)}$ along the interior tree
$T_\inte$. Therefore, the reduced density matrix reads up to a normalisation (later fixed by the
requirement that $\tr_A \rho_\inte =1$):
\be
\rho_\inte(g_{e\in A},\tg_{e\in A})\,=\,\prod_{e\notin T_\inte} \delta(G_e)\delta(\tG_e) \,
\prod_{i=1}^{(V_\rb-1)}\delta(C_i^{-1}C_0\tC_0^{-1}\tC_i).
\ee

The next step is to compute the von Neumann entropy of this density
matrix, $S_A=-\tr_A \rho_\inte
\log \rho_\inte$. We have two types of terms: some $\delta(G_e)\delta(\tG_e)$ for edges not in the tree ${e\notin
T_\inte}$ and some $\delta (g_i \tg_i^{-1})$ which only involve edges in the tree $T_\inte$. Since
these different terms do not involve the same group elements, they can be treated separately (more
precisely, we can do a change of variable in the integrations replacing the $g_e$'s by the $G_e$'s
for the edges not in the tree $T_\inte$ and the Jacobian of the transformation is trivial due to
the left and right invariance of the Haar measure). The terms $\delta(G_e)\delta(\tG_e)$
corresponding to the interior loops are density matrices corresponding to pure states and thus have
a zero entropy. The only non-vanishing contribution therefore comes from the boundary loops and
gives:
\be
S_A= (V_\rb-1) S[\delta(g\tg^{-1})],
\ee
where $S[\delta(g\tg^{-1})]$ is the entropy of the to-be-normalized density matrix
$\sigma(g,\tg)=\delta(g\tg^{-1})$. It turns out that this density matrix $\sigma$ is actually the
identity matrix and its entropy is simply the log of the dimension of the Hilbert space of $L^2$
functions on $\SU(2)$. However, this Hilbert space has an infinite dimension and the result
requires a regularization.

The regularization consists in introducing a cut-off $J$ in the representations of $\SU(2)$. We
define a regularized $\delta_J$ function in term of the $\SU(2)$ characters\footnotemark:
\be
\delta_J(g\tg^{-1})=\sum_{j\in\N/2}^{j\le J}
(2j+1)\chi_j(g\tg^{-1}).
\ee
\footnotetext{
We could also use the $q$-deformation of $\SU(2)$ which provides a natural way to restrict to a
finite number of representations \cite{kassel}. For instance, given the deformation parameter
$q=\exp(i\pi/J)$, the highest representation is $j=J-1/2$. Their $q$-dimensions are equal to
$$
d^{(q)}_j\equiv\f{\sin(2j+1)\pi/J}{\sin\pi/J}=\chi_j(\pi/J),
$$
which is the same as the evaluation of the usual $\SU(2)$ character on an angle $\theta=\pi/J$.
When $J\arr\infty$ (and $q\arr1$), we recover $d^{(q)}_j\arr (2j+1)$. Hence the $q$-deformed
version of the reduced density matrix $\sigma(g,\tg)$ is
$$
\rho_{(q)}(g,\tg)=\sum_{j}d^{(q)}_j \chi_{j}(g\tg^{-1}).
$$
Its entropy (after properly normalization of the density matrix) is
$$
S(\rho_q)=\log\sum_j [d^{(q)}_j]^2= \log N(J,\pi/J)\sim3\log J+\ldots,
$$
where the factor $N(J,\pi/J)$ is introduced later when dealing with topological defects. }
Corresponding we consider the Hilbert space of $L^2$ functions on $\SU(2)$ which decompose onto
matrix elements of the group elements involving only representations $j\le J$. This Hilbert space
$\cH_J$ is spanned by the (renormalized) (Wigner) matrix elements $\sqrt{2j+1}\,D^j_{mn}(g)$, where
$m$ and $n$ run by integer step from $-j$ to $+j$.  Its dimension is thus:
\be
\Delta_J\,\equiv\,\sum_{j\in\N/2}^{j\le J} (2j+1)^2 \,=\, \f13(1+2J)(1+J)(3+4J)\sim
8J^3/3. \label{deltaj}
\ee
It is easy to check that the non-normalized density matrix $\delta_J(g\tg^{-1})$ is the identity on
$\cH_J$ by explicitly computing its matrix elements\footnotemark and therefore its entropy is
$S(\delta_J)=\log\Delta_J$.  Indeed, if we introduce the normalized kets through
$$
\6g|j,m,n\9=\sqrt{2j+1}D^{(j)}(g)^m_{~n},\qquad
\6j,m,n|j',m',n'\9=\delta_{jj'}\delta_{mm'}\delta_{nn'},
$$
then the normalized density matrix
\be
\rho^J=\frac{1}{\Delta_J}\sum_{j\in\N/2}^{j\leq
J}\sum_{m,n}|j,m,n\9\6j,m,n|,
\ee
is obviously maximally mixed, hence is maximally entangled with the rest of the system \cite{hor}.
This finally proves the results which we announced above (a more detailed proof is presented in
Appendix A):
\begin{result}
The entanglement between the region $A$ and the rest of the spin network state for the completely
flat state $\psi_0$ is:
\be
S_A=E_A[\psi_0]\,=\,(V_\rb-1)\,\log \Delta_J,
\qquad\textrm{with}\quad
\log\Delta_J \underset{J\arr\infty}{\sim}3\log J +\log\f83.
\ee
\end{result}
At the end of the day, it is the number of boundary vertices that is relevant for the entanglement.
It could seem surprising since the $\SU(2)$ group elements live on the edges of the graph and
therefore the degrees of freedom of the theory apparently live on the edges. We would then expect
the entanglement to scale with the number boundary edges. However, the requirement of gauge
invariance implies that the physical degrees of freedom actually live on the (independent) loops of
the graph and not simply on its edges. Then we proved that the number of (independent) boundary
loops is directly related to the number of boundary vertices (minus one).
\footnotetext{
We compute, using the orthogonality of $\SU(2)$ matrix elements:
$$
\la\sqrt{2k+1}D^k_{cd}\,|\delta_J|\,\sqrt{2j+1}D^j_{ab}\ra\,=
\int dgd\tg\, \sqrt{(2k+1)(2j+1)} D^j_{ab}(g)\overline{D^k_{cd}}(\tg)\delta_J(g\tg^{-1})
=\delta_{jk} \delta_{ac}\delta_{bd}.
$$ }
A last comment is that we would still have found the same leading order behavior for the
entanglement in case we had first regularize and then calculate the reduced state instead of the
contrary as we did above~\footnotemark.
\footnotetext{
It can be understood by noting that
$$
\int\!dg
\delta(g)\delta(g)=\delta(\1)=\sum_{j\in\N/2}d_j^2\rightarrow\Delta_J,
$$
and regularization of $\delta$-function prior to the integration produces the same result,
$$
\sum_{j,k\in\N/2}^J
d_jd_k\int\!dg\chi_j(g)\chi_k(g)=\sum_{j,k\in\N/2}^Jd_jd_k\delta_{jk}=\Delta_J.
$$}

\subsection{Topological defects and renormalized entanglement}

We can further generalize the procedure described above to take into account topological defects.
In the context of 3d gravity, topological defects represent point particles. For instance, a
(spinless) particle of given mass leads to a non-trivial holonomy around it: we impose
$\delta_\theta(g)$ instead of $\delta(g)$ on loops around it with the angle $\theta$ related to the
mass (see \cite{grav,frei} for more details).

Moreover, we expect that the leading order of the entanglement does
not change when including topological defects. Nevertheless, for any
state $\psi$ we introduce the renormalized (or relative)
entanglement as a regularized entropy \cite{ren} of the reduced
density operator of $A$, $\bar{E}_A[\psi]\equiv\bar{S}(\rho_\inte)$
\be
\bar{S}[\rho_\inte]\equiv\,\lim_{J\rightarrow\infty}S(\rho_\inte^J)-S(\rho_{0\,\inte}^J)
\ee
 which is the difference of the
regularized measure of entanglement of the region $A$ with the rest
computed in the spin network state $\psi$ and the entanglement
computed for the completely flat state $\psi_0$. When $\psi$ only
includes topological defects, we actually expect the renormalized
entanglement to converge to a finite value as the regulator $J$ is
taken to infinity.

Let us first consider inserting a topological defect in an interior loop or an exterior
loop~\footnotemark, i.e replacing $\delta(G_e)$ or $\delta(H_e)$ by $\delta_\theta(G_e)$ or
$\delta_\theta(H_e)$. In the case of an exterior loop, we will get in the reduced density matrix
after integration over the group elements living on the exterior edges a term $\int dH
\delta_\theta(H)^2$ instead of $\int dH \delta(H)^2$.
\footnotetext{
We can further introduce any topological defect on arbitrary products of these interior and
exterior loops of the type $\delta_\theta(G_{e_1}G_{e_2}G_{e_3}..)$ and
$\delta_\theta(H_{e_1}H_{e_2}H_{e_3}..)$. As shown in the detailed proof presented in Appendix A,
this does not affect the entanglement at all, which turns out to depend only on the boundary loops.
}
Such a term only enters the normalization of the reduced density matrix and thus does not affect
the entanglement. In the case of an interior loop, the modified loop constraint gives a term
$\delta_\theta(G_e)\delta_\theta(\tG_e)$ in the reduced density matrix instead of the original
$\delta(G_e)\delta(\tG_e)$. This still represents a pure state, thus has zero entropy. Once again,
it does not affect the entanglement. This proves the following result:
\begin{result}
Starting off from the completely flat state $\psi_0$ and then adding topological defects along some
loops of the spin network state, if no topological defect is inserted on boundary loops then the
entanglement between the region $A$ and the rest of the graph does not change:
\be
\bar{E}_A[\psi]=0, \qquad\textrm{or equivalently}\quad E_A[\psi]=(V_\rb-1)\log\Delta_J.
\ee
\end{result}

On the other hand, if we insert a topological defect on the boundary between $A$ and the outside,
or more precisely replace the constraint $\delta(B_i)$ by  $\delta_\theta(B_i)$ on one given
boundary loop, this will modify our previous entanglement calculation. For the completely flat
state $\psi_0$, the term in $\rho_\inte$ corresponding to that loop was $\int
dgd\tg\,\delta(g\tg^{-1})\delta(g_ig)\delta(\tg_i\tg)=\delta(g_i\tg_i^{-1})$ which gives the
totally mixed state (identity density matrix on the Hilbert space of $L^2$ functions over
$\SU(2)$). This now replaced by:
\be
\int dgd\tg\,\delta(g\tg^{-1}) \delta_\theta(g_ig)\delta_\theta(\tg_i\tg)=
\int dg\, \delta_\theta(g_ig)\delta_\theta(\tg_ig^{-1})=
\sum_j  \f{\chi_j(\theta)^2}{(2j+1)}\chi_j(g_i\tg_i^{-1}),
\ee
where we used the decomposition of the distribution $\delta_\theta$ into $\SU(2)$ representations
as given by eqn.\Ref{deltatheta}. Cutting off the sum over representations to $J$, we compute the
matrix elements of this reduced density matrix:
\be
\la\sqrt{2k+1}D^k_{cd}\,|\sigma_J^{(\theta)}|\,\sqrt{2j+1}D^j_{ab}\ra\,=
\f{1}{N(J,\theta)}\delta_{jk} \left(\f{\chi_j(\theta)}{2j+1}\right)^2\delta_{ac}\delta_{bd},
\ee
where the normalisation factor $N(J,\theta)$ ensures that the reduced density matrix
$\sigma_J^{(\theta)}$ has a unit trace:
\be
N(J,\theta)=\sum_{j\le J}\chi_j(\theta)^2=\frac{(3+4J)\sin\theta-\sin(3+4J)\theta}{4\sin^3\theta}.
\ee
One can check that doing a Taylor expansion around $\theta\sim0$, we recover
$N(J,\theta\arr0)=\Delta_J$. Finally, computing the von Neuman entropy of this density matrix gives
the renormalized entanglement between the region $A$ and the outside in the state $\psi_\theta$
with one topological insertion along a boundary loop:
\begin{result}
Inserting a single topological defect $\delta_\theta$ along a boundary loop between the region $A$
and the outside leads to a non-zero renormalized entanglement:
\be
\bar{E}_A[\theta]
=
\lim_{J\rightarrow \infty}\log \f{N(J,\theta)}{\Delta_J} -\f{1}{N(J,\theta)}\sum_{j\le
J}\chi_j(\theta)^2\log\f{\chi_j(\theta)^2}{(2j+1)^2}.
\ee \label{result3}
\end{result}
We prove in Appendix B that this expression has a finite limit for
all values of $\theta$ when the regulator $J$ is sent to $\infty$.
This renormalized entanglement has a universal value when $\theta$
is {\bf not} a rational fraction of $\pi$:
\be
\bar{E}_A(\theta)=-3+\log 6\approx -1.20824.
\ee
We can also compute this value analytically\footnotemark{} for some specific values of the angle
$\theta$:
$$
\bar{E}_A(\pi/2)=-2+\log(3/2)\approx-1.5945,\quad
\bar{E}_A(\pi/3)=-2+\log 2\approx -1.30436,\quad \bar{E}_A(\pi/4)=-2+\log 3-\half\log2\approx
-1.248.
$$
We see that the renormalized entropy (and entanglement) is discontinuous. This is actually a
generic feature of infinite-dimensional systems \cite{wehrl,rmp}. It is also known that if a
reasonable constraint is put on the accessible set of states, such as bounded mean energy, $\tr
H\rho <\infty$, then the entropy will be a continuous function on this set. In our case the entropy
is continuous if the cut-off $J$ is large but finite. This can be motivated by appealing to the
geometric interpretation of the representation labels. Indeed, in the context of 2+1 (loop)
gravity, the spin network states are the eigenstates of the length operator and its eigenvalues are
exactly the representation labels (up to ordering ambiguities). Thus requiring the finite extent of
the system in space would naturally impose an upper cut-off on the edge representation labels.

\footnotetext{
For example, for $\theta=\pi/2$, $\chi_j(\pi/2)$ vanishes for all half-integer values of $j$ and is
equal to $(-1)^j$ when the representation label $j$ is an integer. We can then compute the
renormalized entanglement using the following exact sum:
$$
\sum_{j\in\N}^{j\le J}\log (2j+1)=-\half\log\pi+\log(2^{J+1}\Gamma(J+\tree)).
$$}

\begin{figure}[htbp]
\epsfxsize=0.45\textwidth
\centerline{\epsffile{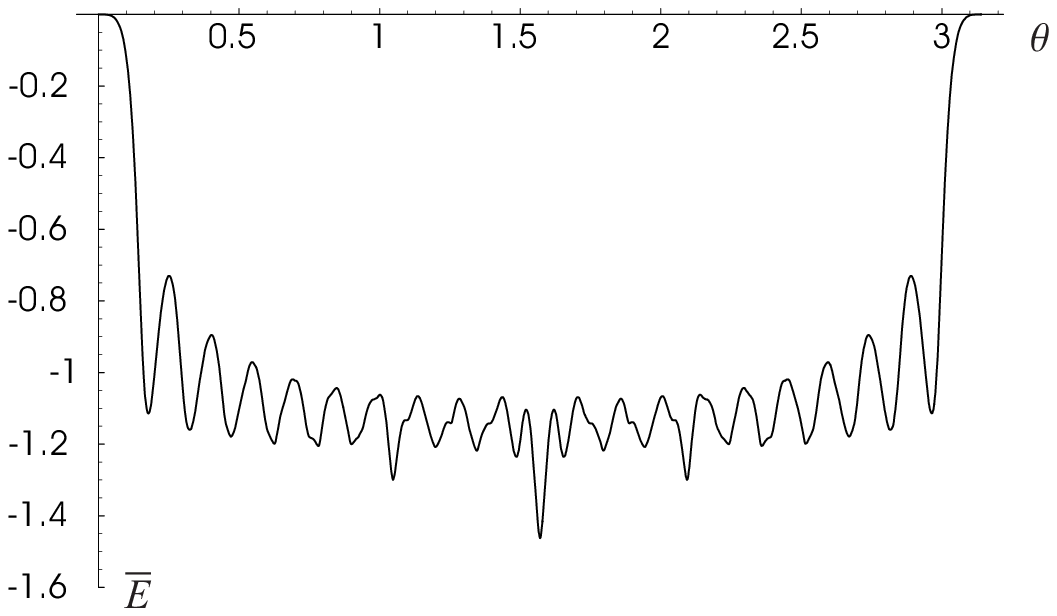}\hspace{0.8cm}\epsfxsize=0.45\textwidth\epsffile{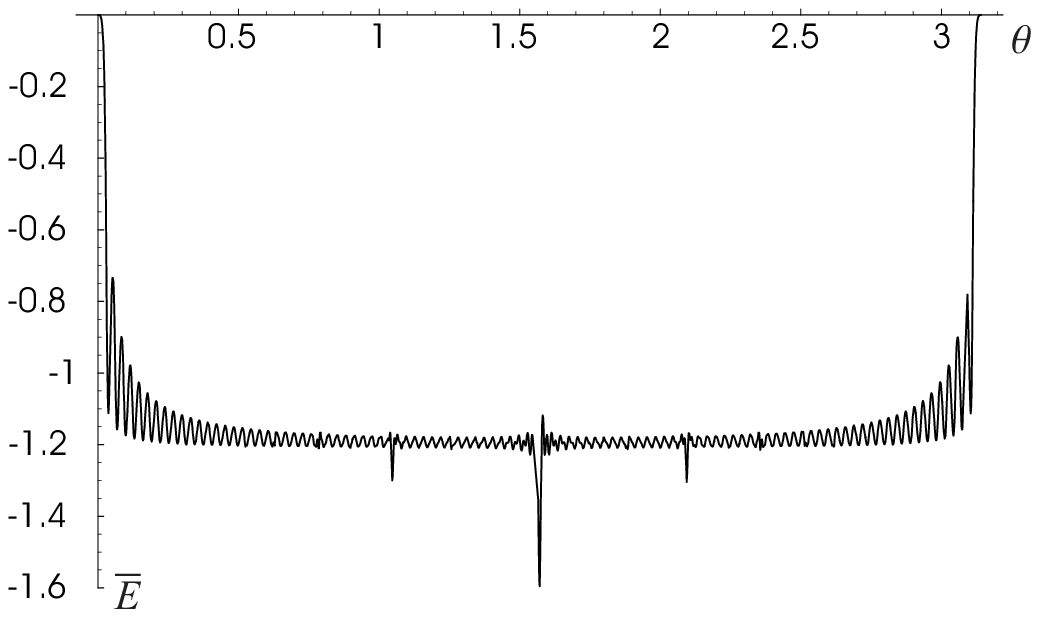}}
\caption{\small{The regularized entanglement $\bar{E}^J[\theta]=E_A^J(\psi_\theta)-E_A^J(\psi_0)$
for different values of the representation cut-off $J=10$ and $J=50\half$}.}\label{entfig}
\end{figure}

The negativity of the above result can be easily understood if we
recall the the flat holonomy state is maximally entangled. Therefor,
the states with non-trivial holonomies are less than maximally
entangled, hence the negativity of the renormalized entropy.

 This result extends to the more general
case of several topological defects inserted along different
boundary loops, each boundary loop contributing independently to the
(renormalized) entanglement between the region $A$ and the rest of
the spin network state.

\medskip

Finally, we are left with the possibility of a non-trivial (2d) topology, i.e non-contractible
cycles in the canonical surface. It is easy to see that the entanglement calculations are not
affected by non-contractible cycles as long as the region $A$ has a trivial topology (isomorphic to
the unit disk). Otherwise, if the region $A$ contains some non-contractible cycles, the holonomies
around them will couple to the holonomies around the cycles outside $A$. As an example, taking the
case of a 2-torus, let us consider that one cycle with holonomy $G$ is contained inside $A$ while
the  dual cycle with holonomy $H$ is outside. Now, flatness of the connection does not require
$\delta(G)\delta(H)$ but the weaker condition that they commute $GHG^{-1}H^{-1}$. Then the reduced
density matrix $\rho_\inte$ for the region $A$ will be given in term of integrals of the type:
$$
\sigma(G,\tG)=\int dH \delta(GHG^{-1}H^{-1})\delta(\tG H\tG^{-1}H^{-1})f(H),
$$
for some central function $f$ (invariant under conjugation). The representation decomposition  of
such density matrix involves the $\{6j\}$ symbol and are more complicated to analyze. We postpone
this study to future investigation.

\section{Some Simple Examples of Entanglement}

\subsection{The $\Theta$-graph} \label{entropy1}

The simplest spin network that allows a non-trivial entropy calculation is the $\Theta$-graph. It
is too simple to be decomposed into two non-trivial inside/outside regions, but it allows to
illustrate how to compute the entanglement between two sets of holonomies.

\begin{figure}[htbp]
\epsfxsize=0.2\textwidth
\centerline{\epsffile{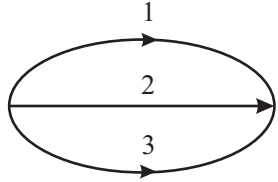}}
\caption{\small The oriented $\Theta$-graph }\label{fig1}
\end{figure}
\noindent
The totally flat state is (the choice of maximal tree is a single edge $T=\{e_3\}$):
\be
\Psi(g_1,g_2,g_3)=\delta(g_1g_3^{-1})\delta(g_2g_3^{-1}),
\ee
where we chose the two independent loops $[e_1,e_3]$ and $[e_2,e_3]$. Notice that the
$\delta$-functions also impose that the loop $[e_1,e_2]$ carries a trivial holonomy,
$g_1g_2^{-1}=\Id$.  Constructing the (formal)  density operator ((up to
regularisation/renormalisation) for the edge $e_3$ one obtains
\be
\rho(g_3,\tilde{g}_3)=\int
dg_1dg_2\delta(g_1g_3^{-1})\delta(g_2g_3^{-1})\delta(g_1\tilde{g}_3^{-1})\delta(g_2\tilde{g}_3^{-1})=
\delta(g_3\tilde{g}_3^{-1})^2=\delta(\1)\delta(g_3\tilde{g}_3^{-1}).\label{formal}
\ee
The expectation values are calculated according to the trace formula $\6O\9=\tr(\rho O)/\tr\rho$.
Hence we regularize the density operator by dropping the infinite multiplicative constant and
truncating the remaining $\delta$-functions in its expansion in term of $\SU(2)$ representations:
\be
\rho^{J}(g_3,\tilde{g}_3)=\frac{1}{N}{\sum_{j_{3}\in\N/2}^{j\leq J}} (2j_3+1)\chi_{j_3}(g_3\tilde{g}_3^{-1})=
\frac{1}{N}{\sum_{j_{3}}}
\sum_{n_3,m_3}(2j_3+1)D^{j_3}_{m_3n_3}(g_3)D^{j_3}_{n_3m_3}(\tg_3^{-1}),
\ee
where the sum on $j_3$ is over all half-integers and the $D^j_{mn}(g)$ are the matrix elements of
the (Wigner) matrix representing the group element $g$ in the representation of spin $j$ (with $m$
and $n$ running by integer step from $-j$ to $+j$).
The normalization constant is determined by the trace condition $\tr\rho=1$, so
\be
N=\Delta_J\equiv\sum_{j\in\N/2}^{j\le J} (2j+1)^2=\frac{1}{3}(1+2J)(1+J)(3+4J)\sim 8J^3/3.
\ee
Introducing the following ket notation (as in \cite{us:geo}),
\be
\6g|j,m,n\9=\sqrt{2j+1}\,D^{j}_{mn}(g),\qquad
\6j,m,n|j',m',n'\9=\delta_{jj'}\delta_{mm'}\delta_{nn'},\label{reduced0}
\ee
the state induced on the edge $e_3$ can be written as
\be
\rho_{3}^{J}=\frac{1}{\Delta_J}\sum_{j_3}\sum_{m_3,n_3}|j_3m_3n_3\9\6j_3m_3n_3|,\label{rho1}
\ee
For a pure state $|\Psi\9$ the von Neumann entropy of $\rho_{(3)}^J$
is the measure of entanglement between the edge $e_3$ and the rest
(see e.g \cite{us:geo}). Hence
\be
E(\Psi|3:1,2):=S(\rho_{3}^J)=\log\Delta_J\sim 3\log J+\log(8/3).
\label{basics}
\ee
The result is obviously divergent when $J\rightarrow\infty$. We can check that computing the
entanglement between the edge $e_1$ and the rest (edges $e_2,e_3$) gives the same result. Moreover,
as a consistency check, we calculate the complementary density matrix $\rho(g_1,\tg_1; g_2,\tg_2)$,
\be
\rho_{(12)}(g_1,  \tilde{g}_1; g_2,  \tilde{g}_2)=\int\! dg_3
\Psi(g_1,g_2,g_3)\Psi(\tg_1,\tg_2,g_3)
=\delta(G_{12})\delta(\tG_{12})\delta(\tilde{g}_1g_1^{-1})
\ee
The loop $G_{12}=g_2g_1^{-1}$ is the analog of an interior loops while $g_1g_3^{-1}$ plays the role
of the boundary holonomy. Finally, as the reduced state $\rho^{(12)}$ is the direct product of the
pure interior state $\delta(G_{12})\delta(\tG_{12})$ and of the mixed state
$\delta(\tilde{g}_1g_1^{-1})$, its entropy is determined only by the latter and we see that
$S(\rho^{(12)}_J)=\log\Delta_J=S(\rho^{(3)}_J)$ as expected.

\medskip

We can further introduce a topological defect along one of the loops of the $\Theta$-graph. Let us
consider the state
\be
\Psi_\theta=\delta_\theta(g_1g_3^{-1})\delta(g_2g_3^{-1}),
\ee
which satisfies the following constraints, for all $j\in\N/2$,
$$
\left|
\begin{array}{l}
\chi_j(g_1g_3^{-1})\Psi_\theta(g_1,g_2,g_3)\,=\,
\chi_j(g_1g_2^{-1})\Psi_\theta(g_1,g_2,g_3)\,=\,
\chi_j(\theta)\Psi_\theta(g_1,g_2,g_3),
\\
\chi_j(g_2g_3^{-1})\Psi_\theta(g_1,g_2,g_3)\,=\,(2j+1)\Psi_\theta(g_1,g_2,g_3).
\end{array}\right.
$$
Expanding the $\delta_\theta$-distribution in $\SU(2)$ characters, we compute the reduced density
matrix for the edge $e_1$:
\be
\rho^{(1)}_\theta(g_1,\tg_1)\,=\,
\sum_{j\in\N/2}\f{\chi_j^2(\theta)}{2j+1}\chi_j(g_1\tilde{g}_1^{-1}).\label{charid}
\ee
After truncation of the sum over representations and proper normalisation, this gives in the ket
notation:
\be
\rho^{(1)}_{\theta}=\f{1}{N(J,\theta)}\sum_{j\le
J}\sum_{m,n}\f{\chi_j^2(\theta)}{(2j+1)^2}|jmn\9\6jmn|,
\qquad
N(J,\theta)\equiv\sum_{j\le
J}\chi_j^2(\theta)=\frac{(3+4J)\sin\theta-\sin(3+4J)\theta}{4\sin^3\theta}.
\ee
For $\theta=0$, we recover the previous reduced density matrix computed for the flat state. This
leads to a regularized entanglement as we obtain in the generic case:
\be
E(\Psi_\theta|1:2,3)=
\log N(J,\theta) -\frac{1}{N(J,\theta)}\sum_{j\le J}\chi_{j}^2(\theta)\log\frac{\chi_{j}^2(\theta)}{(2j+1)^2}.
\ee
On the the hand, we can also compute the reduced density operator $\rho^{(2)}_{\theta}$ for the
edge $e_2$:
\be
\rho^{(2)}_\theta(g_2,\tilde{g}_2)=\int
dg_1dg_3\,\delta_\theta(g_1g_3^{-1})\delta(g_2g_3^{-1})\delta_\theta({g}_1
{g}_3^{-1})\delta(\tilde{g}_2{g}_3^{-1})
\,=\,\delta(g_2\tilde{g}_2^{-1})\int dG\,\delta_\theta(G)^2.
\ee
Up to a normalisation, it is actually equal to the density matrix $\rho^{(2)}$ computed above for
the flat state, $\rho^{(2)}_\theta(g_2,\tg_2) \propto \delta(g_2\tg_2^{-1})=
\rho_{(2)}(g_2,\tg_2)$. Similarly, we obtain that $\rho^{(3)}_\theta(g_3,\tg_3)\propto
\delta(g_3\tg_3^{-1})$.
That means that imposing a non-trivial holonomy around the single loop $[e_1,e_2]$ by
$\delta_\theta(g_1g_2^{-1})$ does not influence the entanglement for the edge $e_3$ vs. $[e_1,e_2]$
or for the edge  $e_2$ vs. $[e_1,e_3]$.

\subsection{Further examples}


\begin{figure}[htbp]
\psfrag{q}{{\large $h$}}
\psfrag{h}{{\large $q$}}
\psfrag{g}{{\large $g$}}
\begin{center}
\includegraphics[width = 4.5cm]{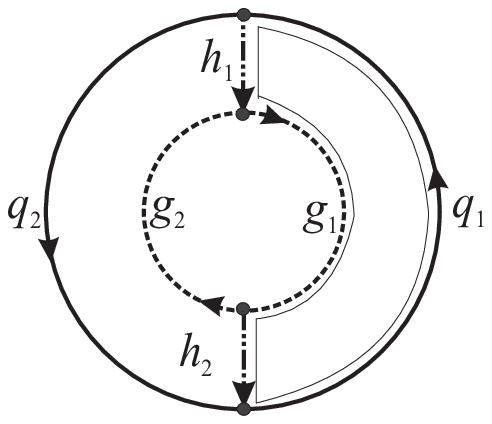}\hspace{0.8cm}\includegraphics[width = 4.5cm]{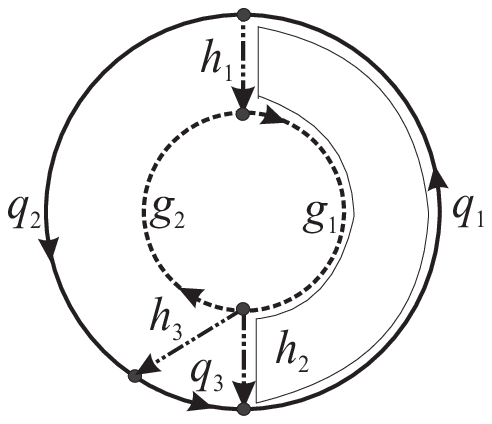}
\caption{\label{wheel} Interior edges are drawn as broken lines, exterior edges are solid. A boundary loop is indicated by
a thin line.}
\end{center}
\end{figure}

A simple graph that is represented on Fig.~\ref{wheel}a has an equal number of all three types of
edges,  $E_\inte=E_\rb=E_\ext=2$ and allows a simple inside/outside distinction. The holonomies
along the internal, boundary and external edges are denoted by $g$, $q$, and $h$, respectively. It
has two internal and two external vertices, so the three independent loops are produced by one
internal, one external, and one boundary loop. The loops are marked on Fig.{\ref{wheel}a. The flat
wave functional is given by (taking into account the orientation of the edges)
\be
\Psi_0=\delta(g_1g_2)\delta(h_1h_2)\delta(g_1q_2h_1q_1).
\ee
Computing the reduced density matrix for the two interior edges gives:
\be
\rho_{\inte}^{\Psi_0}(g,\tg)=\delta(G)\delta(\tG)\delta(\tg_1g_1^{-1}),
\ee
where $G=g_1g_2$ is the holonomy along the interior loop. This interior density matrix decomposes
into a direct product of a pure state on the interior loop and the mixed state
$\delta(\tg_1g_1^{-1})$. According to the previous calculations, the regularized entropy is simply
\be
S(\rho_{\inte}^{\Psi_0})=\log\Delta_J.
\ee
It is easy to check that a different choice of loop or increase in their number only results in the
appearance of additional $\delta(\1)$ factors that do not alter the regularized entropy.

Let us look at the possible addition of a new boundary edge without changing the number of boundary
vertices as in Fig.\ref{wheel}b. The new boundary edge actually does not create a new boundary loop
but simply a new exterior loop (since it does not involve any interior edge). Thus it should not
contribute to the entanglement. Indeed the new flat state is:
\be
\Psi(g,q,h)=\delta(g_1g_2)\delta(h_1h_2h_3)\delta(g_1q_2h_1q_1)\delta(q_3h_3q_2^{-1}).
\ee
As expected, it results in the same reduced density matrix for the interior,
$\rho^{\inte}_{\Psi}=\rho^{\inte}_{\Psi_0}$, thus also
$S(\rho^{\inte}_{\Psi})=S(\rho^{\inte}_{\Psi_0})$.

Coming back to the original graph in Fig.\ref{wheel}a, we impose a non-trivial holonomy around the
boundary loop:
\be
\Psi_\theta=\delta(g_1g_2)\delta(h_1h_2)\delta_\theta(g_1q_2h_1q_1).
\ee
The interior density operator is
\be
\rho_{\inte}^{\Psi_\theta}(g,\tg)=\delta(G)\delta(\tG)\delta_\theta(\tg_1g_1^{-1}),
\ee
and it leads to a renormalized entropy equals to $\bar{E}(\theta)$ as we computed in the previous
section (see result.\ref{result3}).

\section{Conclusions}

We computed the entropy for a bounded region on a physical state of BF theory. Indeed, being
solvable and lacking local degrees of freedom, BF theory allows for explicit calculations and a
precise analysis of the relationship between boundaries and degrees of freedom. Looking at physical
spin network states, we showed that the entanglement between the two regions for an arbitrary
bipartite splitting of the spin network only depends on the structure of the boundary: the entropy
simply scales with the size of the boundary. More precisely, we proved that the entanglement grows
with the number of boundary vertices, $E\propto (V_\rb -1)$. We also showed that the introduction
of topological defects does not affect this result at leading order and we computed exactly the
finite entropy difference due to such particle-like defects on the boundary.

Technically, we developed the necessary mathematical tools required to analyze the graph structure
of a flat distributional spin network state and to regularize the resulting wave functional and
entropy calculations. These tools are also relevant to entanglement calculations for the  spin
systems used for topological quantum computation such as the Kitaev model \cite{kitaev, hamma}.

We hope to apply these techniques to study the precise relation between gauge breaking and entropy
on one hand, and to compute the entanglement for a more general class of spin network states
relevant for loop quantum gravity \cite{future}.

\appendix

\section{Detailed Proof of the Results}

In this Appendix we  discuss  the  structure of independent loops of $\Gamma$ and prove our Results
1--3 in a more general setting. The connected graphs $\Gamma_\inte$ and $\Gamma_\out$ (that
consists of of the outer vertices, $V_\out=V-V_\inte$ and $E_\ext$ exterior edges)
contribute $E_\inte-V_\inte+1$ and $E_\ext-V_\out+1$ independent interior and exterior loops,
respectively. It is convenient to introduce a further classification of the remaining $E_\rb-1$
loops. A true boundary loop in addition to the boundary (and, possibly)
 exterior edges includes at least one interior edge.
Loops that contain at least one boundary edge, but no interior ones, we (for the lack of a better
term) name frontier
 loops.  Exterior loops that were defined above, together with the frontier loops form
all the independent loops of $\Gamma_\ext$ (as defined in Sec. IIA it consists of
$V_\ext=V_\out+V_\rb$ exterior and boundary vertices, and $E_\ext+E_\rb$  exterior and boundary
edges).


 From the definition of
 frontier loops it follows that their number in the set of
independent loops is given by
\be
L_\rf=(E_\ext+E_\rb-V_\out-V_\rb+1)-(E_\ext-V_\out+1)=E_\rb-V_\rb,
\label{frontloop}
\ee
where the total number of the exterior loops is
\be
L_\ext=E_\ext+E_\rb-V_\ext+1
\ee
 Finally, the number of boundary loops
in the set of independent loops of $\Gamma$ is
\be
L_\rb=V_\rb-1.
\ee

 Fig.~\ref{flowers} illustrates the loop construction that is used to define the new variables.
 First we select $L_\inte$ independent interior  loops.
   Next,  $L_\rb=V_\rb-1$
 boundary loops are constructed as follows.
     We select $V_\rb$ different paths ($v_0\, v_i\, w_0$) that link
 the two representative vertices, which are distinguished by the boundary vertices
$v_1,v_2,\ldots, v_{V_\rb}$ they are passing through. The numbering of boundary vertices  induces a
natural numbering of the boundary loops that are formed by combining the sequential paths $(v_0\,
v_{i}\, w_0)$ and $(v_0\, v_{i+1}\, w_0)$. Finally, the set of $L$ independent loops is completed
by a necessary number of the external loops.

 The $i$-th boundary loop carries a holonomy
$B_i\equiv g_i Q_i g_{i+1}^{-1}$, where $g_i$ is the (internal) holonomy acquired on the way from
$v_0$ to $v_i$, and the outer holonomy $Q_i$ is acquired along the path $(v_iw_0 v_{i+1}$). The
holonomies $Q_i$ are the first $L_\rb$ new exterior variables. Holonomies $H_i$ around
$i=1,\ldots,L_\ext$ exterior loops (calculated with respect to $w_0$) provide another $L_\ext$
independent variables. Since
\be
L_\ext+L_\rb=E_\ext+E_\rb-V_\out<E_\ext+E_\rb,
\ee
we complete the set of outer variables by adjoining to the list the holonomy $Q_{V_\rb}$ which is
acquired by going from $v_{V_\rb}$ to $w_0$ along the edge $v_0 v_{V_\rb}w_0$, and picking
additional
 variables $r$.

To express an arbitrary holonomy $R$ as a product of $L$ independent holonomies and their inverses,
all of them should be calculated with respect to some fixed reference point, such as $v_0$. Hence,
we pick up an arbitrary edge (say, $v_{V_\rb}$), so with respect to $v_0$ the outer flower
holonomies become $H_i\rightarrow g_{V_\rb}Q_{V_\rb}H_iQ^{-1}_{V_\rb}g^{-1}_{V_\rb}$.

Holonomies $G_i$, $i=1,\ldots,L_\inte$, around the independent interior loops are part of the set
of new internal variables.
 When $V_\rb=V_\inte$, the
 representative interior point is on the boundary, and can be taken
 to be $v_{V_\rb}$. Then $g_{V_\rb}\equiv\1$, and the set of
 interior variables is completed by the holonomies $g_i$,
 $i=1,{V_\rb}-1$. Indeed, they resulting set is independent and
 \be
 L_\inte+(V_\rb-1)=E_\inte.
 \ee
On the other hand, when $V_\inte>V_\rb$, we add a non-trivial $g_{V_\rb}$ to the set of interior
variables, and possibly more additional variables $\sg$.

In these new variables the flat state $\psi_0$  becomes
\be
\psi_0(g,h,q)=\prod_{r=1}^{L_{\inte}}\delta(G_r)\prod_{s=1}^{L_\rb}\delta(g_sQ_sg_{s+1}^{-1})
\prod_{t=1}^{L_\ext}\delta(H_t)
\prod_i\delta(R_i),
\ee
where the last product contains the delta-functions of additional holonomies. The index $i$ can
cover all possible remaining loops of $\Gamma$. Those holonomies (calculated with respect to $v_0$)
can be expressed as
\be
R_i=(g_{V_\rb}Q_{V_\rb})^{\pi_i}\!\!\prod_{P_i\cdot\{r,s,t\}}\!\!G_r^{\pi_{P_i\cdot
r}}(g_sQ_sg_{s+1}^{-1})^{\pi_{P_i\cdot s}}H_t^{\pi_{P_i\cdot t}}\,\,(g_{V_\rb}Q_{V_\rb})^{-\pi_i},
\ee
where number and ordering of the factors, as well as their powers $\pi_a=0,\pm1$ are determined by
the decomposition of $R_i$. The variable $\tilde{R}_i$ is defined by the same equation with
$g\rightarrow\tilde{g}$, $G\rightarrow\tilde{G}$.

As a result,  the interior density operator of \textbf{Result 1} is
\be
\rho_{\inte}(g,\sg,G;\tilde{g},\tilde{\sg},\tilde{G})=\prod_{r=1}^{L_\inte}\delta(G_r)\delta(\tilde{G_r})\!\int
dQdHdr
\prod_{s=1}^{L_\rb}\delta(g_sQ_sg_{s+1}^{-1})
\delta(\tilde{g}_sQ_s\tilde{g}_{s+1}^{-1})\prod_{t=1}^{L_\ext}\delta^2(H_t)
\prod_i\delta(R_i)\delta(\tilde{R}_i).
\ee
The integration results in
\be
\rho_{\inte}(g,\sg,G;\tilde{g},\tilde{\sg},\tilde{G})=\prod_{r=1}^{L_\inte}\delta(G_r)\delta(\tilde{G_r}))\delta(\1)^{L_\out}
\prod_{s=1}^{L_\rb}\delta(g_{s+1}^{-1}g_s\tilde{g}_s\tilde{g}_{s+1})\prod_i\delta^2(\1),
\label{t1form}
\ee
where the conditions   $g_sQ_sg_{s+1}^{-1}=\1$ and the factors
$\delta(g_{s+1}^{-1}g_s\tilde{g}_s\tilde{g}_{s+1})$ and $\delta(G_r)$, were used to transform $R_i$
to the identity. Dropping the infinite constants and making the final change of variables
$g'_s=g_{s+1}^{-1}g_s$ brings the interior density operator to the form
\be
\rho_{\inte}(g,\sg,G;\tilde{g},\tilde{\sg},\tilde{G})=\sigma_\inte(G,\tilde{G})
\prod_{s=1}^{L_\rb}\delta(g'_s\tilde{g}'^{-1}_s),
\ee
which completes the proof of Result 1.

\textbf{Result 2} deals  with the situation where $L_\rb$ independent boundary loops that carry trivial
holonomies can be chosen.
 In this case the  state is given by
\be
\psi_{\{\theta\}}=
\prod_r^{L_\inte}\delta_{\theta_r}(G_r)
\prod_{s=1}^{L_\rb}\delta(g_sQ_sg_{s+1}^{-1})
\prod_{t=1}^{L_\ext}\delta_{\theta_t}(H_t)
\prod_i\delta_{\theta_i}(R_i),
\ee
for some assignment $\{\theta\}$ to the external and internal independent loops, and any compatible
assignment to the dependent loops.

 After the integration over $Q$ it reduces to
\be
\rho_{\inte}^{\{\theta\}}=\sigma_\inte^{\{\theta\}}(G,\tilde{G})
\prod_{s=1}^{L_\rb}\delta(g_{s+1}^{-1}g_s\tilde{g}_s\tilde{g}_{s+1})\int
dH \delta^2_{\theta_t}(H_t)
\prod_i\delta_{\theta_i}(R'_i)\delta_{\theta_i}(\tilde{R'}_i),
\ee
where $ R'_i=\prod_{P_i\cdot\{r,t\}}\!\!G_r^{\pi_{P_i\cdot r}}H_t^{\pi_{P_i\cdot t}}, $ does not
contain any of the interior variables. Hence the reduced state
\be
\rho_{\inte}^{\{\theta\}}=\sigma^\inte_{\{\theta\}}(G,\tilde{G})f(\theta,G)f(\theta,\tilde{G})
\prod_{s=1}^{L_\rb}\delta(g_{s+1}^{-1}g_s\tilde{g}_s\tilde{g}_{s+1})
\ee
has the same form as in Eq.~(\ref{t1form}). \hfill $\Box$

Finally, in the setting of \textbf{Result 3} we pick the first boundary loop to have a non-trivial
holonomy. Then
\be
\psi_\theta(g,h,q)=\prod_{r=1}^{L_{\inte}}\delta(G_r)\delta_\theta(\tilde{g}_1Q_1\tilde{g}_{2}^{-1})
\prod_{s=2}^{L_\rb}\delta(g_sQ_sg_{s+1}^{-1})
\prod_{t=1}^{L_\ext}\delta(H_t)
\prod_i\delta_{\theta_i}(R_i),
\ee
where $\theta_i$ is either 0,or $\theta$, depending on the loop. Integration over all variables but
$Q_1$ results in the reduced state being given as
\be
\rho_\inte^\theta\propto\sigma_\inte(G,\tilde{G})\prod_{s=2}^{L_\rb}\delta(g_{s+1}^{-1}g_s\tilde{g}_s\tilde{g}_{s+1})
\!\int dQ_1
\delta^{n+1}_\theta(g_1Q_1g_{2}^{-1})\delta^{{n+1}}_\theta(\tilde{g}_1Q_1\tilde{g}_{2}^{-1}),
\ee
where $n$ is a number of ``other" loops that carry holonomies of the class $\theta$. Since
$\delta_\theta^2(g)=\delta_\theta(g)c_\theta$, where the infinite constant can be expressed in a
regularized fashion using the formulas of Ec.~I.A,  the reduced state  finally becomes
\be
\rho_{\inte}^\theta(g,\sg,G;\tilde{g},\tilde{\sg},\tilde{G})=\sigma_\inte(G,\tilde{G})
\delta_\theta(g'_1\tilde{g}'^{-1}_1)\prod_{s=2}^{L_\rb}\delta(g'_s\tilde{g}'^{-1}_s),
\ee
which  trivially exhibits the desired entropy.

\section{Computing the Renormalized Entanglement}

Here we prove that the renormalized entanglement $\bar{E}(\theta)$, as introduced in the Sec.~II.C,
 $$
\bar{E}_J[\theta]=\lim_{J\rightarrow\infty}S(\rho_\inte^{\theta\,J})-S(\rho_{\inte}^J)=\lim_{J\rightarrow\infty}
\log \f{N_J(\theta)}{\Delta_J} -\f{1}{N_J(\theta)}\sum_{j\le
J}\chi_j(\theta)^2\log\f{\chi_j(\theta)^2}{(2j+1)^2},
$$
converges for any value of the angle $\theta$. Because of the symmetry
$\theta\leftrightarrow\pi-\theta$, we can restrict ourselves to $0<\theta\leq\pi/2$. Since
\be
N_J(\theta)\sim \frac{J}{\sin^2\theta},\qquad \Delta_J\sim 8J^3/3,
\ee
it follows that the limit exists if asymptotically
\be
S(\rho_\inte^{\theta\,J})\sim 3\log J+f(\theta)+\cO(1/J),
\ee
for some function $f(\theta)$.

We have to consider two different cases, depending on whether  the class angle is a rational or
irrational fraction of $\pi$. In the first case the class angle is $\theta=p/q\pi$, with relatively
prime $p, q\in\N$, and there are only finitely many different values that $\sin^2n\theta$ takes. It
is periodic with period $K_\theta=q$, and
\be
\sum_{l=0}^q \sin^2(lp\pi/q)=q/2.
\ee
To simplify the notation we further take $J=kq$, $k\in\N$, and consider the limit
$k\rightarrow\infty$.

 As a result, the entropy
\be
S(\rho_\inte^{\theta\,J})\sim\log J-2\log\sin\theta -\frac{1}{J}\sum_{n=0}^{2kq+1}\sin^2
n\theta\log\frac{\sin^2 n\theta}{n^2\sin^2\theta},
\ee
becomes
\be
S(\rho_\inte^{\theta\,J})\sim \log J
+\frac{1}{kq}\sum_{j=0}^{2k}\sum_{l=0}^{q-1}(2\sin^2(l\theta)\log(jq+l)-\sin^2(l\theta)\log[\sin^2(l\theta)])+f(\theta)+\cO(1/J),
\ee
where we expressed the summation index $n$ by $n=jq+l$. The entropy finally reduces to
\be
S(\rho_\inte^{\theta\,J})\sim \log J+\frac{2}{kq}\frac{q}{2}(\log [(2k)!]+2k\log q) +
f(\theta)+\cO(1/J)=3\log J+\tilde{f}(\theta)+\cO(1/J),
\ee
which establishes the claim for $\theta=p/q\pi, p, q\in\N$.

For a generic value of $\theta$ we establish the limit  by  using the Euler--Maclauren integration
formula for sums. The asymptotic behavior of the entropy is
\be
S(\rho_\inte^{\theta\,J})\sim  \log J+\f{1}{J}\left(2\int_0^{2J}\!dn
\sin^2{(n\theta)}\log{n}-\int_0^{2J}\!dn\sin^2{(n\theta)}\log{\sin^2{(n\theta})}\right)+f(\theta)+\cO(1/J).
\ee

Using the known integral $\int_0^\pi dx
\sin^2x\log\sin^2x=\pi(\half-\log 2)$,  the second term becomes
\be
\f{1}{J}\int_0^{2J}\!dn\sin^2{(n\theta)}\log{\sin^2{(n\theta})}\sim 1-2\log 2.
\ee
The first integral has a closed form that involves sine and cosine integral functions, but the
relevant part is simply
\be
\f{1}{J}\int_0^{2J}\!dn \sin^2{(n\theta)}\log{n}\sim -2+2\log2+2\log J+\ldots.
\ee
hence the limit exists for all $\theta$ and for the irrational fractions of $\pi$
\be
\bar{E}_A(\Psi_\theta)=-3+\log 6.
\ee



\end{document}